\documentstyle[12pt,psfig]{article}
\begin{document}

\begin{center}
{\bf PAULI AND ORBITAL EFFECTS OF MAGNETIC FIELD\\ 
ON CHARGE DENSITY WAVES}

\end{center}

\begin{center}
{A. Bjeli\v{s}$^*$, D. Zanchi$^{**}$, and G. Montambaux$^{***}$}
\end{center} 
\begin{center}
{\em \small 
$^*$ Department of Physics, Faculty of Science, University of Zagreb, POB 162, 10001
Zagreb, Croatia \\
$^{**}$ Laboratoire de Physique Th\'eorique et Hautes Energies, 
Universit\'es Paris VI Pierre et Marie Curie -- Paris VII Denis Diderot, 
2 Place Jussieu, 75252 Paris C\'edex 05, France\\
$^{***}$ Laboratoire de Physique des Solides, asssoci\'e au CNRS, 
Universit\'e Paris-Sud, 91405 Orsay, France}
\end{center}

\vspace{6mm}

\begin{quote}
{\small {\bf Abstract.} 
Taking into account both Pauli and orbital effects of external magnetic field 
we compute the mean field phase diagram for charge density waves in 
quasi--one--dimensional electronic systems.
The magnetic field can cause  transitions to CDW states with two types 
of the shifts of wave vector from its zero--field value.
It can also stabilize the field--induced charge density wave.
Furthermore, the critical temperature shows peaks
at a new kind of magic angles.}
\end{quote}

{\bf 1. INTRODUCTORY REMARKS}\\

$\;\;\;\;\;\;$ Although the charge and spin density waves [C(S)DWs] are 
resembling in many physical aspects, they behave differently in the 
external magnetic field. The reason is that the Pauli coupling probes 
different contents of order parameter spaces for two cases. 
The SDW order parameter is of complex vectorial form, in contrast to 
simple complex  parameter space for CDWs. We write four 
complex S/CDW amplitudes in a 
compact form
$
M_i = \Psi^{\dagger}\rho_{+}\sigma_{i}\Psi$, $i=1, ..., 4$,
where  $\Psi^{\dagger}\equiv (\Psi_{\uparrow+}^{\dagger}, 
\Psi_{\uparrow-}^{\dagger}, \Psi_{\downarrow+}^{\dagger}, 
\Psi_{\downarrow-}^{\dagger})$ is the four component fermion field with
$\Psi_{\uparrow(\downarrow)\pm}^{\dagger}$ representing the creation of 
electron  with spin $\uparrow(\downarrow)$ at the right (left)
($k \approx \pm k_F$) quasi-one-dimensional Fermi sheet, and $\rho_{+}$ 
and $\sigma_{i}$ are Pauli matrices in respective ($+, -$) and 
($\uparrow,\downarrow$) two-dimensional spaces, with $\sigma_{4}= I$.
$M_1, M_2$ and $M_3$ are three components of SDW, with $M_3$ chosen along 
the direction of the magnetic field {\bf H}, while $M_4$ is the CDW amplitude. 
The Pauli coupling affects only components 
$M_{3,4}= \Psi_{\uparrow+}^{\dagger}\Psi_{\uparrow-}\pm 
\Psi_{\downarrow+}^{\dagger}\Psi_{\downarrow-}$,
i.e. the linear combinations of two DWs with spins $\uparrow$ and 
$\downarrow$ and with wave numbers $2k_F \pm2q_P$. Here 
$q_P \equiv \mu_B H/v_F$ is the characteristic wave number for the Pauli 
coupling ($\mu_B$ -  Bohr magneton, $v_F$ - longitudinal Fermi velocity). 
Due to this splitting and the mismatch of two Peierls wave numbers, the 
critical temperature for the resulting hybridized combination of $M_3$ and 
$M_4$ is suppressed with respect to its value at {\bf H}= 0.

$\;\;\;\;\;\;$ The systems with the SDW order avoid this suppression by 
orienting its magnetization perpendiculary to the magnetic field. The 
Pauli coupling therefore influences only the spectrum of collective modes. 
It also may introduce a spin flop transition if the magnetic field is 
oriented along the easy axis in systems with internal magnetic
anisotropy.[1] 
The magnetic field influences the SDW phase diagram only through the
another, orbital coupling, characterized by the wave number 
$q_0=ebH \cos \theta$, where b is the transverse lattice constant, 
and $\theta$ is the inclination of the magnetic field with respect to the 
normal to the conducting $(a,b)$ plane. As is well known, this coupling 
has a  general tendency to suppress the effects of imperfect nesting and 
to enhance the critical temperatures.[2]

$\;\;\;\;\;\;$ Since both types of couplings are effective in CDW systems, and the ratio of characteristic scales $\eta \equiv q_0/q_P$ may pass through a wide range of values (particularly by varying the orientation $\theta$), the resulting phase diagrams may well be more complex than that for SDW systems. In the present paper we discuss this problem within a mean  field, random phase approximation (RPA), assuming also only sinusoidal modulations.[3] The extension to non-sinusoidal modulations, but with the Pauli coupling only, was considered elsewhere.[4]

\newpage

{\bf 2. PHASE DIAGRAM}\\

$\;\;\;\;\;\;$The $2 \times 2$ susceptibility matrix for the ($M_3,M_4$) DW response reads[3]
\begin{equation}
\label{mat}
\frac{\chi _g}{f}\left(
\begin{array}{cccc}
\sqrt{1+\delta ^2} + U_c \chi_g          & \delta  \\ 
 \delta                              & \sqrt{1+\delta ^2} + \nu U_c\chi_g
\end{array}
\right)\; ,
\end{equation}
with $\chi_g \equiv \sqrt{\chi_\uparrow \chi_\downarrow}$, 
$\delta \equiv (\chi_\uparrow-\chi_\downarrow)/(2\chi_g)$,
$f \equiv 1 +(1+\nu)U_c\chi_g\sqrt{1+\delta ^2} +\nu U_c^2\chi_g^2$ 
and $\chi_{\uparrow,\downarrow}\equiv\chi_{0}(q_x \pm 2q_P,q_y)$. 
The contributions from the orbital coupling enter in the standard way 
through the bubble polarization diagram 
$\chi_0$.[2,3] The parameter $\nu$ is the ratio of SDW and CDW 
coupling constants,  
$
\nu = U_s/(-U_c) =(2g_1 - g_2)/g_2,
$
with the range of CDW stability defined by $U_c<0,\,\,\nu<1$.
The critical temperature and the wave vector for the CDW in the magnetic field follow from the diagonalization of the matrix (\ref{mat}).

$\;\;\;\;\;\;$ Let us at first assume that the Fermi surfaces
are perfectly nested ($t_b \neq 0, t_{b}^{'}=0$ in the standard notation). 
As the phase diagram in Fig.1(a) shows, at weak enough fields 
[$h\equiv \mu_B H/(2\pi T)\leq h_c=0.304$] the systems keeps the 
perfect nesting wave vector {\bf Q}$_0 = (2k_F,\pi/b)$ (i.e. $q_x=q_y = 0$ 
in eq.\ref{mat}), but the critical temperature $T_{c0}$ decreases with 
respect to its value at $h=0$, 
$T_c^{0} = (2\gamma E_F/\pi)\exp(-\pi v_F/U_c)$. 
As the field passes the critical value $h_c$ 
(that weakly depends on parameters $\eta$ and $\nu$), 
this ordering is replaced by one of those with  shifted wave vectors.
Depending on the subtle relations [3] between the strength of the 
magnetic field, the interaction $U_c$, and the
 the values of parameters  $\eta$ and $\nu$, 
two types of shifts, reflecting qualitatively different effects of Pauli 
coupling, appear possible.

$\;\;\;\;\;\;$ The shift in the longitudinal direction, at some wave numbers
$\pm q_x$ with respect to $2k_F$, is the direct consequence of the 
band Zeeman splitting, present already in the pure one-dimensional limit. 
With $q_x \neq 0$ we have a finite off-diagonal coupling 
of pure CDW and SDW$_z$ in the matrix (\ref{mat}). The resulting  
CDW$_x$ order is therefore a CDW-SDW hybrid, with the relative 
weight which, together with the critical temperature $T_{cx}$ and the 
wave number   $q_x$, depends on the couplings $U_c,\nu$. 
Since CDW$_x$ does not involve 
the orbital coupling, it does not depend on the parameter $\eta$.

$\;\;\;\;\;\;$ Another case is shown in the inset of Fig1(a). 
Between phases CDW$_0$ and CDW$_x$ a phase CDW$_y$ with 
($q_x=0,\,\, \pm q_y \neq 0$) is stabilized. 
It appears for example if we   suppress somewhat  $T_{cx}$ by
introducing a finite negative $\nu$. What remains as maximal critical 
temperature is $T_{cy}$ (itself independent on $\nu$), 
corresponding to the transition metal--CDW$_y$.
CDW$_y$ does not involve SDW$_z$ [since $\delta=0$ in (\ref{mat})], 
but is a pure CDW
with an effective imperfect nesting as a consequence of a finite 
(and large enough) value of $q_P$ in $\chi_g$. The properties of  
CDW$_y$ order follow from the lower diagonal element of the matrix (\ref{mat}).
They are independent on the parameter $\nu$, since SDW channel 
is now not active. On the other hand they depend on the parameter 
$\eta$, since the ``one-dimensionalization'' of the corrugated Fermi 
surface due to the orbital coupling [2] helps in stabilizing this ordering. 
More precisely, the critical temperature $T_y$ is $\eta$-dependent and 
$t_b$-independent, while the wave number $q_y$ and the CDW amplitude at 
$T<T_y$ depend on $t_b$ as well.

$\;\;\;\;\;\;$ In the phase diagram in Fig.1(a) all phase transitions 
($T_{c0}$, $T_{cx}$, and
$T_{cy}$)
between metallic phase and 
three different CDW's are of the second order. The 
transition between CDW$_0$ and $CDW_x$ is of the first order and is
 weaker and weaker (the meta-stability regime gets narrower) as one 
approaches $T_c$ from below. The transition  between CDW$_0$ and $CDW_x$
is at the line $h=0.304$ near  $T_c$, and it deviates at low temperatures 
toward the point $\mu _BH\approx T_c^0$[5], in figure denoted by $P$.
In the inset of Fig.1(a), where intermediate phase CDW$_y$ comes into play,
both transitions $CDW_0$--$CDW_y$ and $CDW_y$--$CDW_x$ are of the first order,
and we expect similar meta-stable regimes around the transitions. 

$\;\;\;\;\;\;$ The perfect nesting--like phase diagram was found in the 
$\alpha-(ET)_2KHg(CSN)_4$ compound[6,7], which indicates that the 
ordered state contains a CDW.
The data are consistent with the case
with one, CDW$_0$--CDW$_x$, transition in magnetic field.
However, it is still not clear to what extent the phase diagram depends on the 
direction of magnetic field.[6,7] The anisotropy of  $T_c$
in magnetic field would indicate that the orbital effects 
are also present, i.e. that the nesting is effectively imperfect.

$\;\;\;\;\;\;$ The phase diagram for imperfectly nested case  
$(t_b'\neq 0)$ is shown in 
Fig.1(b). Note that for $t_b'$ close to its critical value $t_b'^*$ 
(the value of $t_b'$ when $T_c$ completely disappears for $H=0$) the 
phenomenon of field induced CDW appears. The obvious difference with 
respect to FISDW system
is that the Zeeman splitting eventually suppresses  $T_c$.
We expect also a new kind of angular resonances involving the
ratio between orbital and Pauli coupling. They are manifested as  
peaks in $T_c$ at the ``magic angles'' given by
$\eta =2/n$, $n$ being integer. 

\bigskip

\begin{figure}
\centerline{\psfig{file=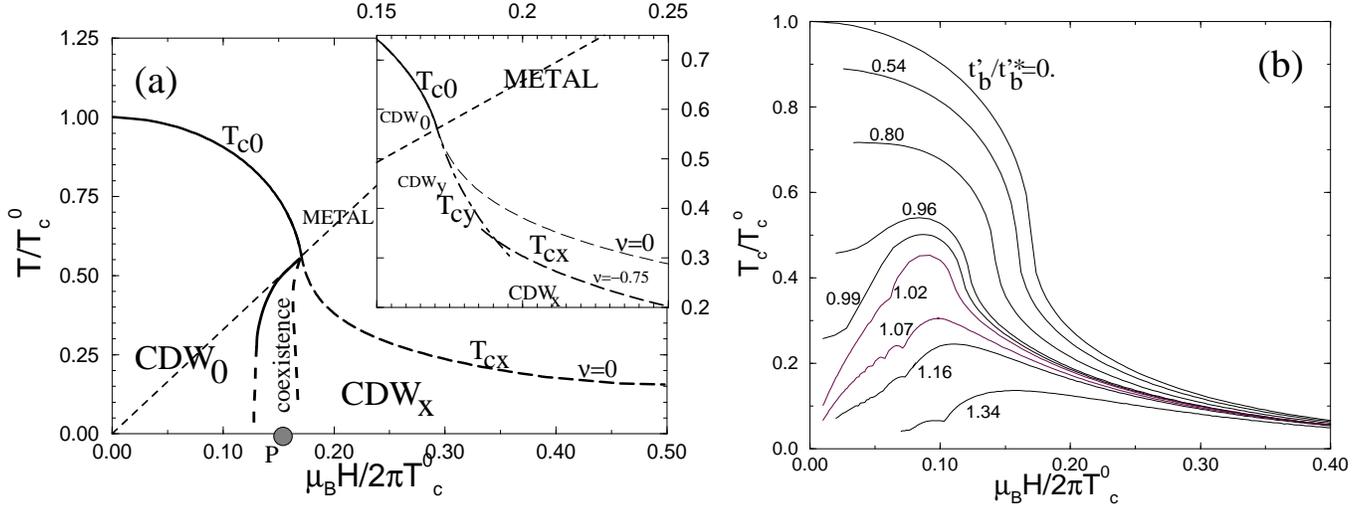,width=18.5cm}}
\caption{ a) The $(H,T)$ phase diagram for the perfect nesting. 
Straight dashed line is given by
$h=\mu_B H/(2\pi T)=0.304$;
Inset shows a case with CDW$_y$  ($\nu =-0.75$).
b)  Set of $T_c$ vs. $H$ curves for different values of imperfect 
nesting parameter.}
\label{Fig}
\end{figure}

{\bf 3. CONCLUSIONS}\\

$\;\;\;\;\;\;$ We have studied the effects of a magnetic field on  
CDWs in  Q1D systems.
Both orbital and Pauli couplings are taken into account in the calculation 
the RPA response in CDW-SDW channel.
At perfect nesting  we explain {\em quantitatively} 
basic features of the phase diagram of $\alpha$-$ET$ compounds.
The field--driven  phase transition [8] and the enhancement 
of the critical 
temperature in magnetic field [9] in  $NbSe_3$ are also
possible experimental realizations of our theory. However, 
due to presumably
 badly nested 
Fermi surface in these materials, the theoretical analysis becomes more
complicated and allow for various possible effects.
More detailed experimental study would be also welcome.
Finally, we emphasize the importance of studying the Q1D compounds of 
MX family in the magnetic field, since there the parameter 
$\nu$ can be tuned by pressure.[10]

\bigskip

{\bf References}\\

\begin{tabular}{rl}
1.& A. Bjeli\v{s} and D. Zanchi, Phys. Rev. {\bf B49},  5968 (1994).\\
2.& L. P. Gor'kov and A. G. Lebed, J. Phys. Lett.(Paris) {\bf 45},
L433 (1984); \\
& M. H\'eritier, G. Montambaux and P. Lederer, J. Physique
Lett. {\bf 45}, L943 (1984).\\
3.& D. Zanchi, A. Bjeli\v{s}, and G. Montambaux, Phys. Rev. {\bf B53}, 
1240 (1996).\\
4.& S. A. Brazovskii and S. I. Matveenko, Zh. Eksp. Teor. Fiz. {\bf
87}, 1400 (1984); \\
& A. Bjeli\v{s} and S. Bari\v{s}i\'{c}, J. Phys. 
{\bf C15}, 5607 (1986).\\
5.& D. Zanchi (unpublished).\\
6.& N. Bi\v{s}kup {\em et al.}, Solid State Commun. {\bf 107}, 503 (1998).\\
7.&  M. V. Kartsovnik {\em et al.}, Synth. Met. {\bf 86}, 1933 (1997); 
JETP {\bf 86}, 578 (1998).\\
8.& P. Monceau and J. Richard,  Phys. Rev. {\bf B37}, 7982
(1988).\\
9.& R. V. Coleman {\em et al.},
Phys. Rev. {\bf B41}, 460 (1990).\\
10.& H. R{\"{o}}der, A. R. Bishop, and J. Tinka Gammel, Synth. Met. {\bf 86}, 2211 (1997).\\

\end{tabular}

\end{document}